\documentclass[aps,prb,twocolumn,groupedaddress,showpacs]{revtex4}

\usepackage{graphicx}
\begin{document}       
\title{Soft x-ray magnetic circular dichroism study of Ca$_{1-x}$Sr$_x$RuO$_3$ across the ferromagnetic quantum phase transition}
\author{J. Okamoto,$^{1,2}$ T. Okane,$^1$ Y. Saitoh,$^1$ K. Terai,$^1$ S.-I. Fujimori,$^1$ Y. Muramatsu,$^1$ K. Yoshii,$^1$ K. Mamiya,$^3$ T. Koide,$^3$ A. Fujimori,$^{1,4}$ Z. Fang,$^5$ Y. Takeda,$^6$ and M. Takano$^7$}
\affiliation{$^1$SRRC, Japan Atomic Energy Research Institute, SPring-8, Sayo-gun, Hyogo 679-5148, Japan \\
$^2$National Synchrotron Radiation Research Center, Hsinchu 30077, Taiwan \\
$^3$Photon Factory, IMSS, High Energy Accelerator Research Organization, Tsukuba, Ibaraki 305-0801, Japan  \\
$^4$Department of Physics, Graduate School of Science, The University of Tokyo, Bunkyo-ku, Tokyo 113-0033, Japan \\
$^5$Institute of Physics, Chinese Academy of Science, Beijing 100080, China \\
$^6$Department of Chemistry, Faculty of Engineering, Mie University, Tsu, Mie 514-0008, Japan \\
$^7$Institute for Chemical Research, Kyoto University, Uji, Kyoto 611-0011, Japan}
\date{\today}
\begin{abstract}
Ca$_{1-x}$Sr$_x$RuO$_3$, which is ferromagnetic for Sr concentration $x$ $>$ 0.3, has been studied by x-ray magnetic circular dichroism (XMCD) in Ru $3p$ and O $1s$ core-level x-ray absorption. XMCD signals appear at $x$ $\sim$ 0.3 and monotonically increases with $x$ in the ferromagnetic phase. While the monotonic increase of the XMCD signals with $x$ is of a typical Stoner-type, the absence of appreciable change in the spectral line shapes of both the Ru 3$p$ and O 1$s$ XMCD spectra indicate that the itinerant-electron ferromagnetism in Ca$_{1-x}$Sr$_x$RuO$_3$ is influenced by strong electron correlation.
\end{abstract}
\maketitle

Recently a number of novel unconventional superconductors in the vicinity of magnetic phases have been discovered, such as UGe$_2$,\cite{UGe2} URhGe\cite{URhGe} and ZrZn$_2$.\cite{ZrZn2} Although the pairing mechanism has not been established in these materials, it is presumed that quantum critical fluctuations are involved for the coexisting/competing ferromagnetism and superconductivity. Magnetic quantum critical transitions have also been observed in Ru oxides. The single-layer Sr$_2$RuO$_4$ is a spin-triplet superconductor with quasi-two-dimensional Fermi liquid state,\cite{Maeno} while non-Fermi-liquid behavior appears in Sr$_2$Ru$_{1-x}$Ti$_x$O$_4$ in the vicinity of antiferromagnetic ordering for the critical impurity concentration of $x_c$$\simeq$ 0.0025.\cite{Sr2RuTiO4} The bilayer perovskite Sr$_3$Ru$_2$O$_7$ shows metamagnetism for a field of $\sim$ 5.5 T below 10 K, and near the metamagnetic field the resistivity shows non-Fermi-liquid bahavior at low temperature.\cite{Sr3Ru2O7}

SrRuO$_3$, the $n$=$\infty$ member of the Ruddelsden-Popper type Ru oxides Sr$_{n+1}$Ru$_n$O$_{3n+1}$, is one of few ferromagnetic metallic oxides ($T_C$ $\sim$ 160 K),\cite{Callanghan,Longo} and its unique ferromagnetism has fascinated many researchers for several decades. Neuimeier $et$ $al$.\cite{Neumeier} have shown that the Curie temperature decreases under hydrostatic pressure. The Rhodes-Wohlfarth ratio $\mu_{eff}/\mu_{ord}$ $\sim$ 1.3 for SrRuO$_3$ is similar to Ni metal, indicating that the magnetic properties of SrRuO$_3$ are close to those of localized electron systems.\cite{Fukunaga} According to photoemission studies of SrRuO$_3$, electron-correlation effects in the Ru 4$d$ bands are relatively strong.\cite{Fujioka,Okamoto} Recent optical studies have shown that SrRuO$_3$ is strongly deviated from a conventional Fermi liquid.\cite{Kostic,Ahn,Dodge1,Dodge2} With substitution of Ca for Sr, the Curie temperature decreases and a ferromagnetic-to-paramagnetic transition occurs at the Sr concentration of $x$ $\sim$ 0.3.\cite{Fukunaga} With Ca substitution, the Ru-O-Ru bond angle decreases from $\sim$ 165$^{\circ}$ to 150$^{\circ}$ but no change in the Ru-O distance has been observed,\cite{Kobayashi} which means that the Ca substitution decreases the Ru 4$d$ band width and that electron correlation within the Ru $4d$ band is enhanced. CaRuO$_3$ is also metallic but does not show long-range magnetic order down to 4.2 K.\cite{Gibbs} At high temperatures, it shows a negative Weiss temperature, suggesting antiferromagnetic correlations. Mukuda $et$ $al$., however, using NMR deduced that CaRuO$_3$ is close to a ferromagnetic metal since the Stoner factor was estimated to be close to 1.\cite{Mukuda} He and Cava have reported that ferromagnetic interaction is observed by replacing Ru by nonmagnetic Ti (Ti$^{4+}$ has electron configuration 3$d^0$) by as small amount as 2 \%.\cite{Cava} Therefore, CaRuO$_3$ is considered to be a metal close to a ferromagnetic instability, and ferromagnetic transition in Ca$_{1-x}$Sr$_x$RuO$_3$ is expected to be a paramagnetic to ferromagnetic quantum phase transition.

Recently, Park $et$ $al$.\cite{Park} and Takizawa $et$ $al$.\cite{Takizawa} have studied the electronic structures of Ca$_{1-x}$Sr$_x$RuO$_3$ using epitaxial thin films by photoemission and x-ray absorption spectroscopy and confirmed that electron-correlation effects increase in going from SrRuO$_3$ to CaRuO$_3$ due to the spectral weight transfer from the coherent to incoherent parts of the Ru 4$d$-band spectra. However, the relationship between the systematic change in the electron correlation strength and the change in the magnetic properties is not clear. In this paper, in order to gain further information about the magnetic properties of the system, a series of Ca$_{1-x}$Sr$_x$RuO$_3$ samples (0$\leq$$x$$\leq$1) have been studied by soft x-ray magnetic circular dichroism (XMCD) in core-level soft x-ray absorption spectroscopy (XAS). 
\begin{figure}
\includegraphics[width=8 cm]{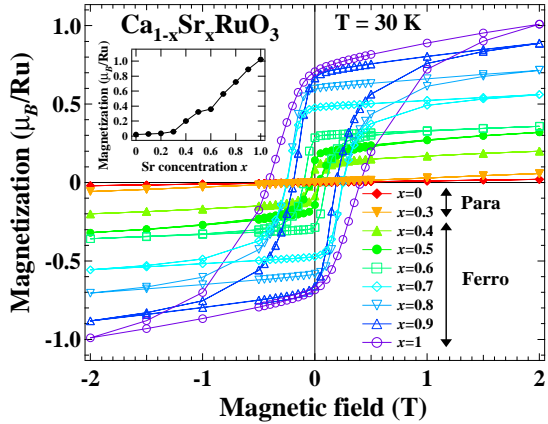}
\caption{\label{SQUID} (Color Online) Magnetization curves of Ca$_{1-x}$Sr$_x$RuO$_3$ (0$\leq$$x$$\leq$1) measured using a SQUID magnetometer at 30 K. The inset shows the magnetization at 2 T as a function of $x$.}
\end{figure}

Sintered polycrystalline samples of Ca$_{1-x}$Sr$_x$RuO$_3$ were prepared in the following procedure. A stoichiometric mixture of RuO$_2$, SrCO$_3$ and CaCO$_3$ powders was prefired at 1000 $^{\circ}$C for 24 hours in air. Then the mixture was pressed into a pellet and fired at 1200 $^{\circ}$C for 48 hours in air. The product was milled, pressed into a pellet again ($\sim$ 2000 Kg/cm$^2$) and fired again at 1400 $^{\circ}$C for 48 hours in air. XAS and XMCD spectra from the Ru $3p$ and O $1s$ core levels were measured at the soft x-ray beamline BL23-SU of SPring-8 in the total-electron-yield mode. The energy resolution was $\sim$ 130 meV at $h\nu$ $\sim$ 700 eV and the degree of circular polarization was estimated to be $\geq$ 95 $\%$ at the Ni $L_{2,3}$ edge from comparison with Ni 2$p$ XMCD reported in the literature.\cite{Chen} The base pressure of the measurement chamber was 1$\times 10^{-8}$ Pa. A fresh surface was obtained before each series of measurements by scraping the samples under an ultra-high vacuum at 30 K. XMCD was measured by switching the helicity of the incident circularly polarized light under a fixed applied magntic field of 2 T. Then, the magnetic field was inverted and the two XMCD spectra were averaged to eliminate suspicious signals. Magnetization of Ca$_{1-x}$Sr$_x$RuO$_3$ were measured using a superconducting quantum interference device (SQUID) magnetometer. 

Figure \ref{SQUID} shows the magnetization curves of the Ca$_{1-x}$Sr$_x$RuO$_3$ samples (0$\leq$$x$$\leq$1) measured at 30 K. Hystereses loops were observed for $x$$\geq$ 0.4, i.e., in the ferromagnetic phase. The magnetization was not saturated up to 2 T, which can be attributed to the increase of the exchange splitting of the Ru $4d$ $t_{2g}$ band under the high magnetic field. Even though hystereses loops were not observed for $x$$\leq$ 0.3, the magnetization at 2 T increases gradually with Sr concentration as shown in the inset of Fig. \ref{SQUID}. Above the critical concentration of $x$$\sim$ 0.3, the magnetization at 2 T increases strongly compared with that in the paramagnetic phase of $x$ $\leq$ 0.3. Coercive force also increases with the Sr concentration for $x$ $\geq$ 0.4. This is in accordance with the scenario that ferromagnetic transition in Ca$_{1-x}$Sr$_x$RuO$_3$ is a paramagnetic to ferromagnetic transition and that the ferromagnetic interaction increases with Sr concentration in the ferromagnetic phase.

\begin{figure}
\includegraphics[width=8 cm]{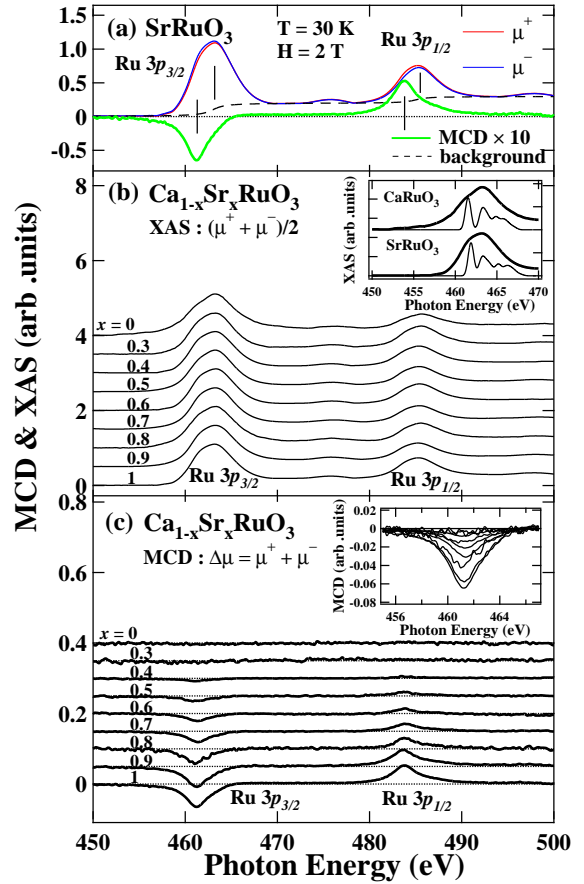}
\caption{\label{RuMCD} (Color online) Ru 3$p$ XAS spectra $\mu^+$ and $\mu^-$ and XMCD spectra $\Delta \mu = \mu^+ - \mu^-$ of Ca$_{1-x}$Sr$_x$RuO$_3$ (0$\leq$x$\leq$1). (a) XAS and XMCD spectra of SrRuO$_3$. Vertical lines correspond to the position of XAS and XMCD peaks. The broken line shows the background of XAS spectrum. (b) Ru 3$p$ XAS [($\mu^+ + \mu^-$)/2] spectra. In the inset of (b), the Ru 3$p_{3/2}$ XAS spectra of CaRuO$_3$ and SrRuO$_3$ are compared with the Ru 4$d$ partial density of states broadened with a Gaussian function (of 1 eV FWHM). (c) Ru 3$p$ XMCD spectra. All the spectra have been normalized to the Ru 3$p_{3/2}$ XAS peak height. The horizontal lines in (c) show zero levels of the XMCD spectra.}
\end{figure}
Figure \ref{RuMCD}(a) shows the Ru 3$p$ XAS spectra for photon helicity parallel ($\mu^{+}$) and antiparallel ($\mu^{-}$) to the Ru 4$d$ majority-spin direction and the XMCD spectrum $\Delta \mu = \mu^+ - \mu^-$ of SrRuO$_3$. The absorption peaks at $\sim$ 463 eV and $\sim$ 485 eV are due to transitions from the Ru 3$p_{3/2}$ and 3$p_{1/2}$ core levels into the Ru 4$d$ band. Other structures located around $\sim$ 476 eV and $\sim$ 498 eV are attributed to transitions into Ru 5$s$ states. The background is due to transition to the continuum above the Fermi level and is composed of two step functions at the Ru $3p_{3/2}$ (463 eV) and $3p_{1/2}$ (485 eV) XAS peaks with the intensity proportional to the degeneracy of the core level. The Ru 3$p$ XAS [($\mu^+ + \mu^-$)/2] spectra and XMCD spectra of Ca$_{1-x}$Sr$_x$RuO$_3$ (0$\leq$$x$$\leq$1) are shown in Figs. \ref{RuMCD} (b) and (c), respectively. In both Ru 3$p_{3/2}$ and 3$p_{1/2}$ XAS spectra, not only peaks but also shoulder structures were observed on the lower photon energy sides of the peaks. In the inset of Fig. \ref{RuMCD}(b), the Ru 3$p_{3/2}$ XAS spectra of CaRuO$_3$ and SrRuO$_3$ are compared with the Ru 4$d$ partial density of states of CaRuO$_3$ and SrRuO$_3$.\cite{calc1} They have been broadened with a Gaussian function of 1 eV FWHM and shifted by 461.3 eV for CaRuO$_3$ and 461.7 eV for SrRuO$_3$. We attribute these peaks and shoulders to the transition to the Ru 4$d$ $e_g$ and $t_{2g}$ bands, respectively. The fact that Ru 3$p$ XMCD signals were only observed for the shoulder structures is consistent with this assignment. The XMCD signal showed symmetric peaks and no appreciable change was observed in the whole Sr concentration range. As for the Ru 3$p$ XAS spectra, their line shape changes throughout the whole Sr concentration range. The energy separation between the shoulder structure and the peak position for CaRuO$_3$ seems to be larger than that for SrRuO$_3$ and this separation gradually becomes smaller as the Sr concentration increases. This agrees with the result of the band-structure calculations\cite{Mazin} that the tilting of the RuO$_6$ octahedra increases the energy separation between the antibonding Ru 4$d$ $t_{2g}$ and $e_g$ bands. In addition, as the Sr concentration increases, the intensity of the shoulder structure increases. This well corresponds to the observation that in the Ru 3$p$ XAS of Ca$_{1-x}$Sr$_x$RuO$_3$ the density of states near the Fermi level increases with Sr concentration.\cite{Lin}

\begin{figure}
\includegraphics[width=8 cm]{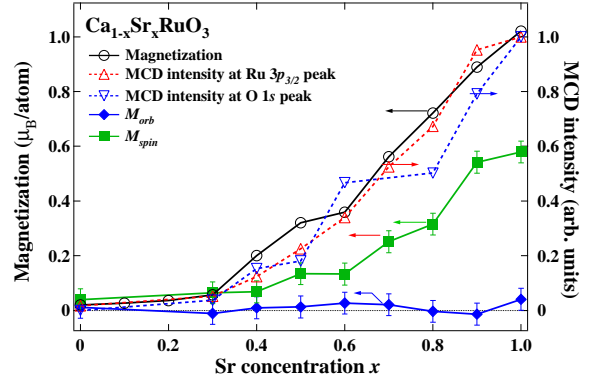}
\caption{\label{SumRule} (Color online) Orbital and spin magnetic moments of the Ru 4$d$ states estimated from the Ru 3$p$ XMCD compared with the magnetization measurements at 30 K under 2 T.}
\end{figure}

For $x$ $>$ 0.3 XMCD signal increased in proportion to the increased Sr concentration as shown in the inset of Fig. \ref{RuMCD}(c) and in Fig. \ref{SumRule}. This is consistent with the above observation (Fig. \ref{SQUID}) that the magnetization in the Ru 4$d$ band increased linearly as the Sr concentrantion $x$ increased beyond the critical value of $x$ $\sim$ 0.3. Such a behavior is different from the simple Stoner model of itinerant ferromagnetism, according to which changes in the line shape of the XMCD spectra are expected to occur across the ferromagnetic transtion due to the exchange splitting of the Ru 4$d$ band.

\begin{figure}
\includegraphics[width=8 cm]{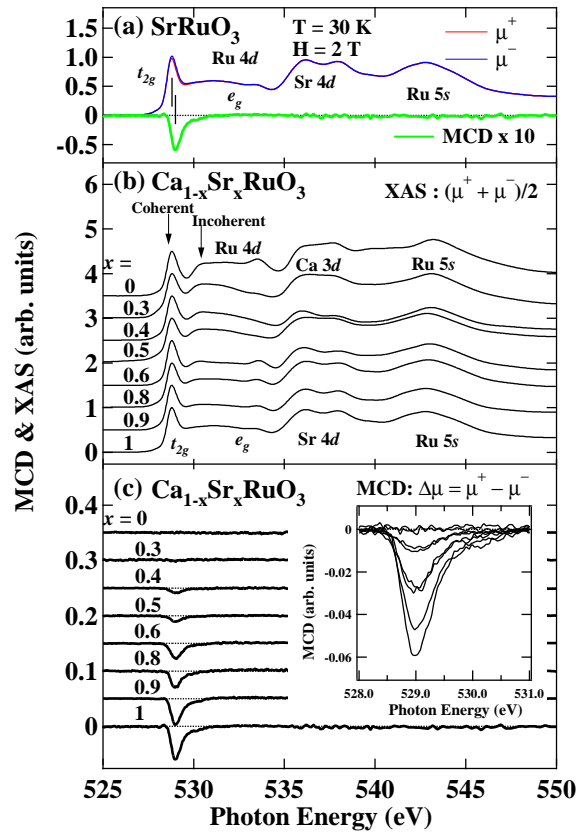}
\caption{\label{OKXAS} (Color online) O 1$s$ XAS spectra $\mu^+$ and $\mu^-$ and XMCD spectra $\Delta \mu = \mu^+ - \mu^-$ of Ca$_{1-x}$Sr$_x$RuO$_3$ (0$\leq$x$\leq$1). (a) XAS and XMCD of SrRuO$_3$. Vertical lines correspond to the position of XAS and XMCD peaks. (b) O 1$s$ XAS spectra [($\mu^+ + \mu^-$)/2]. (c) O 1$s$ XMCD spectra. All the spectra have been normalized to the O 1$s$ XAS peak height at $\sim$ 529 eV. The horizontal lines in (c) show the zero levels of the XMCD spectra.}
\end{figure}
In order to investigate the influence of the magnetization on the ligand O $2p$ states, XMCD measurements were also made in O 1$s$ XAS. Figure \ref{OKXAS} shows the O 1$s$ XAS and XMCD spectra of Ca$_{1-x}$Sr$_x$RuO$_3$. The O 1$s$ XAS spectra represent the unoccupied part of the Ru 4$d$ band mixed with O $2p$ orbitals: Transitions into the Ru 4$d$ $t_{2g}$ and $e_g$ bands are in the regions $528-530.5$ eV and $530.5-534$ eV, respectively.\cite{Imp} The peak at $\sim$ 529 eV and the shoulder structure around 530 eV correspond to the coherent and incoherent parts of the Ru 4$d$ $t_{2g}$ bands, respectively, as reported by Takizawa $et$ $al$.,\cite{Takizawa} indicating strong electron correlation in the Ru $4d$ $t_{2g}$ band. The Sr 4$d$ and Ca 3$d$ states are in the region $534-540$ eV, and the Ru $5s$ states are in the region $540-550$ eV. Significantly, a large negative XMCD structure was observed in the Ru 4$d$ $t_{2g}$ band region, indicating that the influence of the magnetization is strong only in the Ru $4d$ $t_{2g}$ band region and negligibly small in the other regions. The XMCD signal became the largest at $x$ = 1 amounting to $\sim$ 7 \% of the O 1$s$ peak intensity.

Figure \ref{OKXAS}(b) shows that as the Sr concentration $x$ increases, the intensity of O 1$s$ XAS decreases in the incoherent part and increases in the coherent part. Since the number of electrons in the Ru 4$d$ band does not change with $x$, this means that spctral weight is transferred from the incoherent part to the coherent part within the Ru 4$d$ $t_{2g}$ band, i.e., electron correlation strength decreases within the Ru 4$d$ $t_{2g}$ band. As for the O $1s$ XMCD spectra shown in Fig. \ref{OKXAS}(c), the XMCD intensity increases linearly as the Sr concentration $x$ increases in the ferromagnetic phase ($x \geq 0.3$) without any appreciable change in the spectral line shape. Since the energy-integrated intensity of the O 1$s$ XMCD spectrum is proportional to the orbital magnetic moment of the O 2$p$ states,\cite{Thole1} the orbital magnetic moment of the O 2$p$ states increases linearly in the ferromagnetic phase.

By applying the orbital\cite{Thole1} and spin sum rules\cite{Carra1} to the Ru 3$p$ XMCD spectra, we can estimate the orbital and spin magnetic moments of the Ru 4$d$ states. In the case of compounds, especially, transition-metal oxides, however, magnetic moments estimated by the XMCD sum rules tend to be smaller than the magnetization measured by SQUID.\cite{LSCO,SFCO} One of the cause of the discrepancy is the mixing of $p_{3/2}$ and $p_{1/2}$ components. In order to compensate the underestimation of the spin magnetic moment due to the electron--core-hole interaction, which mixes $p_{3/2}$ and $p_{1/2}$ components, correction factors were calculated.\cite{Teramura} Although these correction factors were calculated only for the 3$d$ transition-metal ions, expected correction factor for the Ru 3$p$ XMCD would be smaller and cannot explain this discrepancy, because the influence of the electron--core-hole interaction decreases along the transition-metal series with increasing spin-orbit splitting\cite{Teramura,Schwitalla} and the spin-orbit splitting in the Ru 3$p$ core level is as large as $\sim$ 20 V.

Another possible cause of the discrepancy is that the magnetization of the O $2p$ electrons is also substantial. The magnetic moment of the O $2p$ orbitals is induced by charge transfer from the O $2p$ orbitals to the spin-polarized Ru 4$d$, however, there is no method to estimate the spin magnetic moment of the O $2p$ electrons. Another possible cause is that the magnetization is reduced at the surface compared to that in the bulk. Since the effective probing depth of the XAS measurement is at most 10 nm in the total electron yield mode, the estimated magnetization may be substantially influenced by electronic states at the surface. The estimated magnetization may increase by using cleaved samples or single-crystalline film samples.\cite{Terai} Finally, the use of the XMCD sum rules is strictly valid for atomic wave functions and may not be quantitatively applicable to the itinerant Ru 4$d$ electrons. 

We have estimated the orbital ($M_{orb}$) and spin ($M_{spin}$) magnetic moments of the Ru 4$d$ states using the XMCD sum rules as follows:
\begin{equation}
M_{orb} = -2\frac{\Delta A_{M_3} + \Delta A_{M_2}}{3(A_{M_3}+A_{M_2})}(10 -N_{4d}),
\label{Morb}
\end{equation}
\begin{equation}
M_{spin} + 7M_T= -\frac{\Delta A_{M_3} - 2\Delta A_{M_2}}{A_{M_3}+A_{M_2}}(10 -N_{4d}),
\label{Mspin}
\end{equation}
where $M_{orb}$, $M_{spin}$ and the magnetic-dipole moment $M_T$ are given in units of $\mu_{B}$/atom, $N_{4d}$ is the 4$d$ electron occupation number which is assumed to be 4, $\Delta A_{M_3}$ and $\Delta A_{M_2}$  are the energy integrals of the 3$p_{3/2}$ and 3$p_{1/2}$ XMCD intensities, and $A_{M_3}$ and $A_{M_2}$ are the energy integrals of the 3$p_{3/2}$ and 3$p_{1/2}$ XAS intensities, respectively. In estimating the XAS intensities, background has been subtracted from the XAS specrtra as shown in Fig. \ref{RuMCD}(a). In estimating the spin magnetic moment from the XAS and XMCD spectra, we have to separate the Ru 3$p$ XMCD spectra into the 3$p_{3/2}$ and 3$p_{1/2}$ components. We have divided the Ru 3$p$ spectra into the two components at 478 eV, where the Ru 3$p$ XAS shows a minimum. Since our measurements were made on polycrystalline samples, the angle average would result in a vanishing magnetic-dipole term\cite{Stohr} and therefore we have ignored $M_T$ in estimating the spin magnetic moment using Eq. (\ref{Mspin}).

In Fig. \ref{SumRule}, we compare the XMCD intensities of Ru 3$p_{3/2}$ ($\sim$ 461.5 eV) and O 1$s$ ($\sim$ 529 eV) with the magnetization measured at 2 T and the orbital and spin magnetic moments of Ru 4$d$ estimated from the Ru 3$p$ XMCD spectra. In the paramagnetic phase ($x$ $\leq$ 0.3), however, no orbital magnetic moment is estimated from the Ru 3p XMCD spectra since no XMCD signals are observed in the Ru 3$p$ within experimental error in as shown in Fig. \ref{RuMCD}(c). This indirectly means that the magnetism of Ca$_{1-x}$Sr$_x$RuO$_3$ is strongly influenced by the change of electron correlation through magnetic transition. As for the O 2$p$ states, the O 1$s$ XMCD spectra were relatively sharp but the energy integral of the XMCD intensity was very tiny compared with that of the XAS intensity, leading to nearly the same orbital moment as that of the Ru 4$d$. The orbital magnetic moment of O 2$p$ was estimated to be as small as $\leq$ 1$\times$10$^{-2}$ $\mu_B$/atom in the entire Sr concentration range on the assumption that the number of holes in the O 2$p$ orbitals is $\sim$1. Although the absolute value estimated from XMCD was only $\sim$ 60 $\%$ of the magnetization, the overall $x$ dependence of the XMCD intensities and estimated magnetic moments qualitatively followed the $x$ dependence of the magnetization. 

The XMCD peak intensities and the magnetization increased linearly above $x$ $\sim$ 0.3 as shown in Fig. \ref{SumRule}. These observations are consistent with the Stoner-type ferromagnetism. On the other hand, as no appreciable spectral change in the Ru 3$p$ and O 1$s$ XMCD spectra was observed as a function of $x$, there was no clear indication of the exchange splitting of the Ru 4$d$ band. These observations as well as the large Rhodes-Wohlfarth ratio mean that ferromagnetism in Ca$_{1-x}$Sr$_x$RuO$_3$ is strongly influnced by electron correlation of the Ru $4d$ $t_{2g}$ band and is different from the Stoner-type itinerant ferromagnetism.

The authors would like to thank the staff of BL23-SU for their valuable technical support and M. Takizawa and D. J. Huang for fruitful discussions. This work was partially supported by a Grants-in-Aid for Scientific Research in Priority Area "Invention of Anomalous Quantum Materials" and S17105002 from Japan Society for the Promotion of Science, Japan.


\begin{thebibliography}{99}

\bibitem{UGe2} S. S. Saxena, P. Agarwal, K. Ahilan, F. M. Grosche, R. K. W. Haselwimmer, M. J. Steiner, E. Pugh, I. R. Walker, S. R. Julian, P. Monthoux, G. G. Lonzarich, A. Huxley, I. Sheikin, D. Braithwaite, and J. Flouquet, Nature (London) {\bf 406}, 587 (2000).

\bibitem{URhGe} D. Aoki, A. Huxley, E. Ressouche, D. Braithwaite, J. Flouquet, 
J. P. Brison, E. Lhotel, and Carley Paulsen, Nature (London) {\bf 413}, 613 (2001).

\bibitem{ZrZn2} C. Pfleiderer, M. Uhlarz, S. M. Hayden, R. Vollmer, H. v. L{\"o}hneysen, N. R. Bernhoeft, and G. G. Lonzarichk, Nature (London) {\bf 412}, 58 (2001).

\bibitem{Maeno} Y. Maeno, H. Hashimoto, K. Yoshida, S. Nishizaki, T. Fujita, J. G. Bednortz, and F. Lichtenberg, Nature (London) {\bf 372}, 532 (1994).

\bibitem{Sr2RuTiO4} N. Kikugawa and Y. Maeno, Phys. Rev. Lett. {\bf 89}, 117001, (2002).

\bibitem{Sr3Ru2O7} R. S. Perry, L. M. Galvin, S. A. Grigera, L. Capogna, A. J. Schofield, A. P. Mackenzie, M. Chiao, S. R. Julian, S. I. Ikeda, S. Nakatsuji, Y. Maeno, and C. Pfleiderer, Phys. Rev. Lett. {\bf 86}, 2661 (2001).




\bibitem{Callanghan} A. Callanghan, C. W. Moeller and R. Ward, Inorg. Chem. {\bf 5}, 1572 (1966).

\bibitem{Longo} J. M. Longo, P. M. Raccah and J. B. Goodenough, J. Appl. Phys. {\bf 39}, 1327 (1968).

\bibitem{Neumeier} J. J. Neumeier, A. L. Cornelius, and J. S. Schilling, Physica B {\bf 198}, 324 (1994).


\bibitem{Fukunaga} F. Fukunaga and N. Tsuda, J. Phys. Soc. Jpn. {\bf 63}, 3798 (1994).


\bibitem{Fujioka} K. Fujioka, J. Okamoto, T. Mizokawa, A. Fujimori, I. Hase, M. Abbate, H. J. Lin, C. T. Chen, Y. Takeda, and M. Takano, Phys. Rev. B {\bf 56}, 6380 (1997).

\bibitem{Okamoto} J. Okamoto, T. Mizokawa, A. Fujimori, I. Hase, M. Nohara, H. Takagi, Y. Takeda, and M. Takano, Phys. Rev. B 60, 2281 (1999).

\bibitem{Kostic} P. Kostic, Y. Okada, N.C. Collins, Z. Schlesinger, J.W. Reiner, L. Klein, A. Kapitulnik, T.H. Geballe, and M.R. Beasley, Phys. Rev. Lett. {\bf 81}, 2498 (1998).

\bibitem{Ahn} J.-S. Ahn, J. Bak, H.S. Choi, T.W. Noh, J.E. Han, Yunkyu Bang, J.H. Cho, and Q.X. Jia, Phys. Rev. Lett. {\bf 82}, 5321 (1999).

\bibitem{Dodge1} J.S. Dodge, E. Kulatov, L. Klein, C.H. Ahn, J.W. Reiner, L. Mieville, T.H. Geballe, M.R. Beasley, A. Kapitulnik, H. Ohta, Yu. Uspenskii, and S. Halilov, Phys. Rev. B {\bf 60}, R6987 (1999).

\bibitem{Dodge2} J.S. Dodge, C.P. Weber, J. Corson, J. Orenstein, Z. Schlesinger, J.W. Reiner, and M.R. Beasley, Phys. Rev. Lett. {\bf 85}, 4932 (2000).

\bibitem{Kobayashi} H. Kobayashi, M. Nagata, R. Kanno, and Y. Kawamoto, Mat. Res. Bull. {\bf 29}, 1271 (1994).

\bibitem{Gibbs} T. Gibbs, R. Greatrex, N. N. Greenwood, D. C. Puxley, and K. G. Snowden, J. Solid State Chem. {\bf 11}, 17 (1974).

\bibitem{Mukuda} H. Mukuda, K. Ishida, Y. Kitaoka, K. Asayama, R. Kanno, and M. Takano, Phys. Rev. B {\bf 60}, 12279 (1999).

\bibitem{Cava} T. He and R. J. Cava, Phys. Rev. B {\bf 63}, 172403 (2001).

\bibitem{Park} J. Park, S.-J. Oh, J.-H. Park, D. M. Kim, and C.-B. Eom, Phys. Rev. B {\bf 69} 85108 (2004).

\bibitem{Takizawa} M. Takizawa, D. Toyota, H. Wadati, A. Chikamatsu, H. Kumigashira, A. Fujimori, M. Oshima, Z. Fang, M. Lippmaa, M. Kawasaki, and H. Koinuma, 
Phys. Rev. B {\bf 72}, 060404 (2005).

\bibitem{Chen} C. T. Chen, N. V. Smith and F. Sette, Phys. Rev. B {\bf 43}, R6785 (1991).



\bibitem{calc1} The calculation have been done using the generalized gradient approximation method.

\bibitem{Mazin} I. I. Mazin and D. J. Singh, Phys. Rev. B {\bf 56}, 2556 (1997).

\bibitem{Lin} B. N. Lin, C. Y. Lin, Y. S. Wu, and H. C. Ku, J. Mag. Mag. Mat. {\bf 272-276}, 479 (2004).

\bibitem{Imp} In the O 1$s$ XAS spectra of Ca$_{1-x}$Sr$_x$RuO$_3$ in Fig. \ref{OKXAS}, the dip structure at $\sim$ 534 eV may be due to the contamination state owing to carbon oxides. Because of little influence on the region mixed with the Ru 4$d$ states we ignore it in the following discussion.

\bibitem{Thole1} B. T. Thole, P. Carra, F. Sette, and G. van der Laan, Phys. Rev. Lett. {\bf 68}, 1943 (1992).

\bibitem{Carra1} P. Carra, B. T. Thole, M. Altarelli, and X. Wang, Phys. Rev. Lett. {\bf 70}, 694 (1993).

\bibitem{LSCO} J. Okamoto, H. Miyauchi, T. Sekine, T. Shidara, T. Koide,
 K. Amemiya, A. Fujimori, T. Saitoh, A. Tanaka, Y. Takeda, and M. Takano, 
Phys. Rev. B {\bf 62}, 4455 (2000).

\bibitem{SFCO} J. Okamoto, K. Mamiya, S.-I. Fujimori, T. Okane, Y. Saitoh, 
Y. Muramatsu, K. Yoshii, A. Fujimori, A. Tanaka, M. Abbate, T. Koide, 
S. Ishiwata, S. Kawasaki, and M. Takano, Phys. Rev. B {\bf 71}, 104401 (2005).

\bibitem{Teramura} Y. Teramura, A. Tanaka, and T. Jo, J. Phys. Soc. Jpn. {\bf 65}, 1053 (1995).

\bibitem{Schwitalla} J. Schwitalla and H. Ebert, Phys. Rev. Lett. {\bf 80}, 4586 (1998).

\bibitem{Terai} In our preliminary study of a single-crystalline thin film of SrRuO$_3$, XMCD signal in Ru $3p$ XAS became stronger than the present polycrystalline work by $\sim$ 10 \%.

\bibitem{Stohr} J. St{\"o}hr and H. K{\"o}nig, Phys. Rev. Lett. {\bf 75}, 3748 (1995).



\end{thebibliography}
\end{document}